\documentclass[12pt]{article}
\usepackage[author={Baldur Magnusson}]{pdfcomment}
\usepackage{amsmath} 
\usepackage{color, soul}
\usepackage[rgb]{xcolor}
\usepackage{graphicx,psfrag, epsf}
\usepackage{enumerate}
\usepackage{natbib}
\usepackage{booktabs}
\usepackage{multirow}
\usepackage{bm}
\usepackage{url} 
\usepackage{soul}
\usepackage{tabulary}
\RequirePackage[noblocks]{authblk}
\RequirePackage{hyperref} 
\RequirePackage{amssymb}

\RequirePackage{fullpage}

\usepackage{sectsty}
\allsectionsfont{\sffamily}

\def\eqsecnum{
    \@newctr{equation}[section]
    \def\theequation{\hbox{\normalsize\arabic{section}-\arabic{equation}}}}

\DeclareMathOperator{\logit}{logit}
\DeclareMathOperator{\expit}{expit}
\DeclareMathOperator{\pri}{\mathcal I}
\DeclareMathOperator{\prb}{\mathcal B}
\DeclareMathOperator{\prd}{\mathcal D}
\DeclareMathOperator{\prh}{\mathcal H}

\begin{document}

\def\spacingset#1{\renewcommand{\baselinestretch}%
{#1}\small\normalsize} \spacingset{1}


  \title{\bf Bayesian inference for a principal stratum estimand to assess the treatment effect in a subgroup characterized by post-randomization events}
  \author{Baldur P.~Magnusson}
	
	\author{Heinz Schmidli}
	\author{Nicolas Rouyrre}
\affil{Biostatistics \& Pharmacometrics, Novartis Pharma AG, Basel, Switzerland}
	\author{Daniel O.~Scharfstein}
	\affil{Department of Biostatistics, Johns Hopkins Bloomberg School of Public Health, Baltimore, MD USA}
  \maketitle


\bigskip
\begin{abstract}
The treatment effect in a specific subgroup is often of interest in randomized clinical trials. 
When the subgroup is characterized by the absence of certain post-randomization events, a naive analysis on the subset of patients without these events may be misleading. The principal stratification framework allows one to define an appropriate causal estimand in such settings. 
Statistical inference for the principal stratum estimand hinges on scientifically justified assumptions, which can be included with Bayesian methods through prior distributions. Our motivating example is a large randomized placebo-controlled trial of siponimod in patients with secondary progressive multiple sclerosis. The primary objective of this trial was to demonstrate the efficacy of siponimod relative to placebo in delaying disability progression for the whole study population. However, the treatment effect in the subgroup of patients who would not relapse during the trial is relevant from both a scientific and regulatory perspective. Assessing this subgroup treatment effect is challenging as there is strong evidence that siponimod reduces relapses. Aligned with the draft regulatory guidance ICH E9(R1), we describe in detail the scientific question of interest, the principal stratum estimand, the corresponding analysis method for binary endpoints and sensitivity analyses. 
\end{abstract}

\noindent%


\section{Introduction} \label{sec:intro}

Assessments of the treatment effect in specific subgroups often guide decisions regarding product labeling, reimbursement and clinical practice. 
\cite{yusuf1991} distinguish between two types of subgroups in randomized clinical trials: ``proper subgroups" characterized by baseline data, and ``improper subgroups" characterized by post-randomization data. 
For ``improper subgroups", a naive analysis would compare test and control treatment in patients observed to fall into the subgroup.
Such analyses may be misleading as post-randomization data are potentially affected by treatment.
Nevertheless, subgroups characterized by post-randomization data continue to be of scientific, regulatory and practical interest \citep{hirji2009,bohula2015,ridker2018}.

In a seminal paper, \cite{frangakis2002} proposed principal stratification to overcome the issues with ``improper subgroups".
In this framework, questions related to such subgroups are expressed in the language of potential outcomes \citep{neyman1923,rubin1974}.
For each patient, we envisage his/her potential post-randomization data when randomized to test and control treatment, respectively.
The principal strata then consist of those patients which would fall into subgroups defined by different combinations of their potential outcomes.
As potential outcomes can be seen as baseline covariates, the principal stratum is a ``proper subgroup", and hence the treatment effect in this subgroup, the {\it principal stratum effect}, has a causal interpretation.

Statistical inference is challenging as the principal stratum effect is not fully identifiable from the observed data without additional assumptions. In the following, we use a Bayesian model-based approach for which partial identifiability causes no real difficulty \citep{lindley1972bayesian}. In the Bayesian approach, substantive assumptions are transparently introduced through priors on model parameters, which can be changed in sensitivity analyses. A tutorial on principal stratification is provided by \cite{egleston2010}, and recent research includes \cite{Baccini2017}, \cite{ding2017}, and \cite{egleston2017}. 

Principal stratification is one of five strategies described in the draft regulatory guidance document ICH E9(R1) on estimands and sensitivity analyses in clinical trials \citep{E9:2017}.
The approach is also discussed in the \cite{NAC2010} report commissioned by the FDA.
However, experience with the principal stratification framework in a regulatory context is currently very limited.
In the following, we describe a case study where this framework was used to address a key regulatory question regarding product labeling.
This should facilitate further innovative applications of this framework by practitioners.

Our motivating example is the EXPAND study, a large placebo-controlled trial of siponimod in patients with secondary progressive multiple sclerosis (NCT01665144 on ClinicalTrials.gov).
The primary objective of this trial was to demonstrate the efficacy of siponimod relative to placebo in delaying disability progression, which was achieved \citep{Kappos2018}.
However, the treatment effect in the subgroup of patients who would not relapse during the trial is relevant from both a scientific and regulatory perspective. Assessing this subgroup treatment effect is challenging as there is strong evidence that siponimod reduces relapses \citep{Selmaj2013,Kappos2018}. 

In Section \ref{sec:motivate} we discuss in more detail the EXPAND study and the scientific question of interest.
The principal stratum estimand and scientific assumptions for identification are described in Section \ref{sec:ps}.
Section \ref{sec:model} provides the Bayesian analysis model and methods for sensitivity analysis.
Case study results are then shown in Section \ref{sec:case}. 
The article closes with a discussion (Section \ref{sec:discuss}).
Software code for the main analysis is provided as supplementary information. 

\section{Motivating case study - the EXPAND trial} \label{sec:motivate}

EXPAND was a randomized, double-blind, placebo-controlled, event- and exposure-driven phase 3 study evaluating the efficacy of siponimod in patients with secondary progressive multiple sclerosis (SPMS). 1651 patients were randomized in a 2:1 ratio to receive once-daily 2mg siponimod or placebo. The primary objective was to demonstrate efficacy of siponimod relative to placebo in delaying the time to 3-month confirmed disability progression (CDP) as measured by the Expanded Disability Status Scale (EDSS). The EDSS is an ordinal scale used for assessing neurologic impairment in MS based on a neurological examination. It combines scores in seven functional systems and an ambulation score, and ranges from 0 (no impairment) to 10 (death due to MS). 3-month CDP is defined by a pre-specified increase from EDSS baseline that is subsequently sustained for at least 3 months. The study achieved its objective with an estimated hazard ratio of 0.79 (95\% CI: 0.65--0.95). Full study results are reported in \cite{Kappos2018}.

While relapses in SPMS are relatively infrequent compared to the preceding relapsing-remitting disease stage, some patients did experience relapses during the study. During these relapses, patients experience an increased EDSS score from which they may fully or partly recover (a CDP confirmation may not take place during a relapse). As expected based on prior trials \citep{Selmaj2013}, siponimod also substantially reduced the annualized relapse rate with a rate ratio of 0.45 (95\% CI: 0.34--0.59). This raised the question of siponimod's ability to delay CDP unrelated to its effect on relapses. Of particular interest was the effect of siponimod among the subgroup of patients for whom relapses would be absent during the study. 

Special care must be taken when defining a treatment effect through such a subgroup, since naive classification based simply on absence of on-study relapses would constitute an ``improper" subgroup, defined by post-randomization outcomes that are affected by treatment.  To overcome this issue, a principal stratum estimand was defined as the relative effect of treatment on the occurrence of 3-month CDP within a time interval from time 0 (randomization) to $t^*$ (e.g.~2 years) among the subgroup of EXPAND patients who would not relapse between randomization and $t^*$  regardless of treatment assignment. 
The critical distinction between this estimand and one based purely on observed occurrence of on-study relapse is that the principal stratum estimand is defined in terms of potential outcomes and is hence a proper subgroup.  

\section{Principal stratum estimand} \label{sec:ps}

In this section, we provide a mathematical representation of the estimand defined in Section \ref{sec:motivate} and discuss the extent to which it can be identified from the observed data.

\subsection{Principal stratification}

The principal strata are defined in terms of potential outcomes, see for example 
\citep{neyman1923,rubin1974}. Let $Z=0,1$ be an indicator variable denoting treatment assignment, 
with 0 and 1 corresponding to control and active treatment, respectively. For this 
exposition, we consider a fixed time-interval of interest from time $t=0$ 
(randomization) to time $t=t^*$, e.g.~12 or 24 months after randomization. Now, define
for each subject the following binary potential outcomes:
\begin{itemize}
\item $S(z) = $ occurrence of relapse over the period $[0,t^*]$ 
under treatment $z=0,1$;
\item $Y(z) = $ occurrence of CDP over the period 
$[0,t^*]$ under treatment $z=0,1$.
\end{itemize}
Note that
all subjects have a pair of potential outcomes, only one of which is observed 
(i.e.~the potential outcome corresponding to the treatment actually assigned). Also 
note that in a randomized trial, the potential outcomes are independent of the 
treatment assignment. Finally, we denote by $S$ and $Y$ the corresponding {\it observed} outcomes for the intercurrent event and the primary outcome, respectively. 

We now define four principal strata based on the potential outcomes $S(z)$:
\begin{itemize}
\item {\bf Immune ($\pri$):} Subjects that would not experience relapse regardless of treatment assignment, i.e.~$S(0)=S(1) =0$;
\item {\bf Doomed ($\prd$):} Subjects that would experience relapse regardless of treatment assignment, i.e.~$S(0)=S(1)=1$;
\item {\bf Benefiter ($\prb$):} Subjects that would only experience relapse if treated with control, i.e.~$S(0)=1, S(1)=0$;
\item {\bf Harmed ($\prh$):} Subjects that would only experience relapse if treated with the active treatment, i.e.~$S(0) = 0, S(1)=1$.
\end{itemize} 


As discussed above, randomization ensures that the potential outcomes $S(0)$ and 
$S(1)$ are jointly independent of treatment assignment, i.e.~$\{S(0), S(1)\} \perp Z$.
In other words, the value of the pair $\{S(0), S(1)\}$ is not affected by treatment 
and can thus be regarded as a pre-treatment characteristic. 

We are now ready to define the estimand of interest:
\begin{equation}\label{eqn:estimand}
\frac{P\left[Y(1) =1 | S(0) = 0, S(1)=0\right] }{ P\left[Y(0) = 1 | S(0) = 0, S(1) = 0\right]} = \frac{P[Y(1)=1 | \pri]}{P[Y(0)=1|\pri]}.
\end{equation}
This is the principal stratum causal effect of treatment on CDP (expressed as a risk ratio) in the population of immune patients.

We consider inference about the estimand in Equation (\ref{eqn:estimand}) using the Bayesian framework; see Section \ref{sec:model} for the model specification. One could also proceed with estimation using the frequentist framework, although doing so would likely require several identifying assumptions in order to produce a point estimate (or one could resort to estimating bounds). It is nevertheless illuminating to consider the extent to which the estimand is identifiable. We hence discuss this topic in Section \ref{sec:identification}.

\subsection{Assumptions} 

We make two main assumptions:

\noindent {\bf Assumption 1 -- Joint Exchangeability:} $Z \perp \{S(1), S(0), Y(1), Y(0)\}$. That is, the treatment assignment is jointly independent of the potential outcomes. In our setting, the trial is randomized so this assumption holds by design. 

\noindent {\bf Assumption 2 -- Monotonicity:} For any patient, $S(0) = 0 \Rightarrow S(1) = 0$ or, equivalently, $S(1) = 1 \Rightarrow S(0)=1$.

The monotonicity assumption rules out the ``harmed'' principal stratum and allows some patients to be classified as belonging to exactly one stratum. For example, a patient on the control arm ($Z=0$) with $S=0$ must be ``immune'', and a patient on the active arm ($Z=1$) with $S=1$ must be ``doomed''. 

The monotonocity assumption warrants careful consideration. It is a partially testable assumption as it implies that $P(S(0)=0) \leq P(S(1) = 0)$ and these probabilities are identified under Assumption 1. In fact, we can rule out Assumption 2 if $P(S(0)=0) > P(S(1) = 0)$. However, failure to rule out Assumption 2 does not mean that it is true. Thus, one must ultimately justify the assumption based on substantive considerations. Analyses that assess sensitivity to departures from monotonicity may be warranted. We discuss options for such sensitivity analyses in Section \ref{sec:model}. 

\subsection{Estimand identification}\label{sec:identification}

In this section we discuss identifiability of the estimand defined in Equation (\ref{eqn:estimand}). First, assumptions 1 and 2 allow identification of the principal strata proportions as follows:
\begin{align*}
P[\prd] = P\left[\textrm{Doomed}\right] &= P\left[S(1) = 1, S(0) = 1\right] & \\
&= P\left[S(1) = 1\right] \qquad &(\textrm{monotonitity}) \\
&= P\left[S(1)=1 | Z=1\right] \qquad &(\textrm{exchangeability}) \\
&= P\left[S=1 | Z=1\right].
\end{align*} 
The last equation follows from the {\it consistency} assumption that is fundamental to causal inference, see for example \cite{hr2018}. Hence, the proportion of doomed patients can be estimated by the proportion of patients experiencing the event $S$ in the active arm. Similarly, the proportion of immune patients is estimated by $P(\pri) = P(S=0 | Z=0)$. Finally, the proportion of benefiters is simply obtained by $P(\prb) = 1- P(\pri) - P(\prd)$. 

Now, we briefly discuss the extent to which the estimand of interest can be identified from the data. By applying the conditional probability formula, followed by the monotonicity and exchangeability assumptions, the denominator of Equation (\ref{eqn:estimand}) is identified as follows:
\begin{align*}
P\left[Y(0) = 1 | \pri\right] &= P\left[Y= 1 | S = 0, Z = 0\right]
\end{align*}
Hence the denominator is estimated as the proportion of outcomes $Y$ among control patients for whom $S=0$ in the period of interest. However, the numerator of Equation (\ref{eqn:estimand}) is not identifiable because patients for which $S(1) = 0$ could belong to either the immune stratum or the benefiter stratum. We will hence derive bounds on the numerator which lead to a range of feasible values for the estimand of interest. 

We can use the law of total probability (making no further assumptions) to write
\[
P\left[Y(1) = 1 | \pri \textrm{ or } \prb\right] = P\left[Y(1) = 1 | \pri\right] P\left[\pri | \pri \textrm{ or } \prb\right] + P\left[Y(1) = 1 | \prb\right] P\left[\prb | \pri \textrm{ or } \prb\right]
\]
which can be re-arranged as
\begin{equation}\label{eqn:numerator}
P\left[Y(1) = 1 | \pri\right] = \frac{P\left[Y(1) = 1 | \pri \textrm{or} \prb\right]}
{P\left[\pri | \pri \textrm{ or } \prb\right]} 
- \frac{P\left[\prb | \pri \textrm{ or } \prb\right]}{P\left[\pri | \pri \textrm{ or } \prb\right]} P\left[Y(1) =1 | \prb\right]
\end{equation}
On the right-hand side of this equation, all quantities except $P[Y(1) = 1 | \prb]$ are identifiable from the observed data. Specifically, $P\left[Y(1) = 1 | \pri \textrm{or} \prb\right] = P[Y=1 | S=0, Z=1]$, and $P\left[\prb | \pri \textrm{ or } \prb\right]$ and $P\left[\prb | \pri \textrm{ or } \prb\right]$ are functions of the identifiable principal strata proportions. When we evaluate the right-hand side over the theoretical range of values for $P[Y(1) = 1 | \prb]$ (i.e.~0 to 1), we obtain the range of feasible values for $P[Y(1)=1 | \pri]$. If $\pri$ is large relative to $\prb$ (i.e. $P[\prb | \pri\textrm{ or }\prb]/P[\pri | \pri\textrm{ or } \prb]$ is small) then the range of feasible values will be narrow; conversely if $\prb$ is large relative to $\pri$ then this range will be wide. 

In order to further identify the numerator, additional substantive assumptions would be required. In some settings, for example, it might be reasonable to assume that $P[Y(1) = 1 | \prb] \leq P[Y(1) = 1 | \prd]$. This untestable assumption, which says that probability of the outcome under treatment is lower in the benefiter stratum than the doomed stratum, would narrow the range further but will not lead to full identifiability. Full identifiability will ultimately require several additional untestable assumptions which may call into question the reliability of any conclusions drawn from such an analysis. Rather than imposing additional assumptions to obtain full identifiability, we will use the Bayesian framework to draw inference about the target estimand. 

\section{Modeling and inference} \label{sec:model}

\subsection{Statistical model} \label{sec:bayes}

Let $G$ be a random variable indicating principal stratum membership and define $\pi_g = P(G=g)$ to be the probability of belonging to stratum $g\in \{\pri, \prd, \prb, \prh\}$. Note that we include the harmed ($\prh$) stratum in the model specification; the monotonicity assumption is realized with the use of a strongly informative prior. 

Next, define $\theta_g(z) = \logit\big\{P\left[Y(z) = 1 | G= g\right]\big\}$ as the stratum-specific probability (on the logit scale) of the primary outcome under treatment $z$. The estimand of interest is then expressed mathematically as $\expit\left[\theta_{\pri}(1)\right]/\expit\left[\theta_{\pri}(0)\right]$.

Let $\bm\omega$ be a vector of all model parameters ($\pi_g$ and $\theta_g(z)$). Working in the Bayesian framework, the posterior distribution of $\omega$ is 
\begin{align}\label{eqn:model}
p(\bm\omega|Y,S,Z) &\propto p(Y,S | Z, \bm\omega) \cdot p(\bm\omega) \nonumber \\ 
&= \underbrace{p(Y|S,Z,\bm\omega)}_{\textrm{disability model}} \cdot \underbrace{p(S|Z,\bm\omega)}_{\textrm{relapse model}} \cdot \underbrace{p(\bm\omega)}_{\textrm{prior}}.
\end{align}
The relapse model is given as
\[
p(S | Z,\bm\omega) = (1-Z) \cdot \textrm{Bernoulli}(\pi_{\prb} + \pi_{\prd}) + Z \cdot \textrm{Bernoulli}(\pi_{\prd} + \pi_{\prh}).
\]
Because each combination of $(S,Z)=(s,z)$ implies membership in one of two principal strata, the disability model for $Y$ given $S=s, Z=z$ is represented by a mixture of two Bernoulli distributions. For example, the distribution of $Y$ given $S=0, Z=1$ is (under consistency and randomization) a mixture of two Bernoullis with respective success probabilities $\expit\left[\theta_{\pri}(1)\right]$ and $\expit\left[\theta_{\prb}(1)\right]$, and mixing proportions $\pi_{\pri}/(\pi_{\pri}+\pi_{\prb})$ and $\pi_{\prb}/(\pi_{\pri}+\pi_{\prb})$. Table \ref{tab:likelihood} spells out the likelihood for every combination of $S$ and $Z$.

\begin{table}
\caption{Implied principal strata (PS) and disability model depending on the observed value for $(S,Z)$. \label{tab:likelihood}}
\centering
\begin{tabular}{|c|c|l|}
\hline
{ $(S,Z)$} & Implied PS & Disability model \\ \hline
(0,0) & $\pri$ or $\prh$ & $\frac{\pi_{\pri}}{\pi_{\pri}+\pi_{\prh}}\textrm{Bernoulli}\Big(\expit\big(\theta_{\pri}(0)\big)\Big) + \frac{\pi_{\prh}}{\pi_{\pri}+\pi_{\prh}}\textrm{Bernoulli}\Big(\expit\big(\theta_{\prh}(0)\big)\Big)$\\
\hline $(0,1)$ & $\pri$ or $\prb$ & $\frac{\pi_{\pri}}{\pi_{\pri}+\pi_{\prb}}\textrm{Bernoulli}\Big(\expit\big(\theta_{\pri}(1)\big)\Big) + \frac{\pi_{\prb}}{\pi_{\pri}+\pi_{\prb}}\textrm{Bernoulli}\Big(\expit\big(\theta_{\prb}(1)\big)\Big)$\\
\hline $(1,0)$ & $\prd$ or $\prb$ & $\frac{\pi_{\prd}}{\pi_{\prd}+\pi_{\prb}}\textrm{Bernoulli}\Big(\expit\big(\theta_{\prd}(0)\big)\Big) + \frac{\pi_{\prb}}{\pi_{\prd}+\pi_{\prb}}\textrm{Bernoulli}\Big(\expit\big(\theta_{\prb}(0)\big)\Big)$\\
\hline $(1,1)$ & $\prd$ or $\prh$ & $\frac{\pi_{\prd}}{\pi_{\prd}+\pi_{\prh}}\textrm{Bernoulli}\Big(\expit\big(\theta_{\prd}(1)\big)\Big) + \frac{\pi_{\prh}}{\pi_{\prd}+\pi_{\prh}}\textrm{Bernoulli}\Big(\expit\big(\theta_{\prh}(1)\big)\Big)$\\ \hline
\end{tabular}
\end{table}

To complete the model specification within the Bayesian framework we need to assign prior distributions for the parameters. For the stratum probabilities we use a log-odds scaled categorical distribution with real valued parameters $\alpha_g$. This transformation allows straightforward incorporation of covariates (see Section \ref{sec:missing}) and improves sampling efficiency \citep{carpenter2017}. The principal stratum probabilities $\pi_g$ are then recovered with the softmax function: 
\[
\pi_g = \textrm{softmax}(\alpha_g) = \frac{\exp(\alpha_g)}{\sum_k \exp(\alpha_k)},\ \alpha_{\prb}=0\ \textrm{for identifiability}.
\]
The softmax function maps the 4-dimensional vector of real numbers $\bm{\alpha}=(\alpha_{\pri},\alpha_{\prd}, \alpha_{\prb},\alpha_{\prh})$ to a 4-dimensional simplex, i.e.~ensures that $\sum_g \pi_g = 1$.  We set $\alpha_{\prb}=0$ because the softmax function is invariant under adding a constant to each component of its input. 

We parameterize the active-arm outcome parameter as $\theta_g(1) = \theta_g(0) + \Delta_g$. We then assume independent weakly informative normal priors for the principal stratum parameters $\alpha_{\pri}$, $\alpha_{\prd}$ and the outcome parameters $\theta_g(0)$ and $\Delta_g$, $g\in\{\pri,\prd,\prb,\prh\}$. The specific priors used in our case study are shown in Section \ref{sec:casepriors}.

The monotonicity assumption is encoded through the use of a strongly informative prior on $\alpha_{\prh}$ with  extreme location (relative to a plausible range of the $\alpha_g$ parameters) and small standard deviation. Use of such a prior will essentially imply that $P(\pi_{\prh}=0) = 1$, and that this probability will remain equal to 1 in the posterior distribution. Because monotonicity is enforced through a strongly informative prior, a natural way to assess sensitivity to the assumption is to gradually relax the prior towards ``weaker" forms of monotonicity. To this end, one could both shift the location of the prior closer to zero and increase the prior standard deviation to be closer to that used for the other principal strata. 

\subsection{Inclusion of covariates and handling of missing data}\label{sec:missing}

Let $\bm X$ denote a vector of baseline covariates and $M$ be an indicator variable that denotes whether $(Y,S)$ is missing, i.e., $M=1$ if $(Y,S)$ is missing and $M=0$ if $(Y,S)$ is observed.  We assume that $M$ is independent of $(Y,G)$ given $\bm X, Z$.  We also note that $Z$ is independent of $(Y(0),Y(1),S(0),S(1),\bm X)$, due to randomization. To include covariates and handle missing data, we further index the model parameters from the previous section by $\bm X$, so that $\pi_{g,\bm x} = P[G=g|\bm X=\bm x]$ and expit$\left[\theta_{g,\bm x}(z)\right] = P[Y(z)=1|G=g,\bm X=\bm x]$.  In (\ref{eqn:model}), the disability and relapse models are conditioned on $\bm X$ and $M=0$, where conditioning on $M=0$ follows from our assumption on the missingness mechanism. Further, $p(S|Z,\bm X,M=0,\bm\omega) = (1-Z)\cdot\textrm{Bernoulli}(\pi_{\prb,\bm X}+\pi_{\prd,\bm X}) + Z\cdot\textrm{Bernoulli}(\pi_{\prd,\bm X}+\pi_{\prh,\bm X})$, and $p(Y|S,Z,\bm X,M=0,\bm\omega)$ is a Bernoulli mixture with parameters indexed by $\bm X$; for example, if $(S,Z)=(0,0)$, we have a mixture with parameters $\expit\left[\theta_{\pri,\bm X}(0)\right]$ and $\expit\left[\theta_{\prh,\bm X}(0)\right]$ with mixture weight $\pi_{\pri,\bm X}/({\pi_{\pri,\bm X}+\pi_{\prh,\bm X}})$. 

To recover the marginal quantities $\pi_g$ and $\theta_g(z)$, we use the following formulae: 
\begin{align}\label{eqn:standardize}
\expit\left[\theta_g(z)\right] &= \frac{1}{\pi_g} \int \expit\left[\theta_{g,\bm x}(z)\right]\pi_{g,\bm x}dF(\bm x), \\
\pi_g &= \int \pi_{g,\bm x}dF(\bm x),  \nonumber
\end{align}
where $dF(\bm x)$ denotes the joint cumulative distribution function of $X$. 

In our case study, we use two dichotomous covariates (see Section \ref{sec:case}). We thus obtain estimates of covariate-specific parameters $\pi_{g,\bm x}$ and $\theta_{g,\bm x}(z)$ by fitting (\ref{eqn:model}) separately within each of the four covariate combinations. In general, if covariates are continuous or have too many categories to be treated independently, models could be fitted with regression techniques. For example, the principal stratum parameters could be estimated with multinomial logistic regression with $\alpha_{g,\bm x} = \bm x'\bm \beta_g$ where $\bm \beta_g$ is a vector of covariate coefficients corresponding to principal stratum $g$. 

\section{The EXPAND trial -- Principal stratum analysis} \label{sec:case}

In the context of the EXPAND trial, the principal stratum of non-relapsers corresponds to the immune stratum defined in Section \ref{sec:ps}. We conducted separate analyses for three difference choices for time $t^* = 12, 18$ and $24$ months. Two covariates were used:~baseline EDSS score dichotomized to high ($6+$) and low ($<6$), and occurrence of relapses within 2 years prior to study (yes/no). Hence four covariate strata were obtained. 

\subsection{Prior distributions}\label{sec:casepriors}

We used the following prior distributions in our analysis: 
\begin{itemize}
\item $\alpha_{g,\bm x}\sim N(0,2^2)$ for $g \in \{\pri, \prd\}$ with the following rationale. Without covariates, a $N(0,1)$ prior for $\alpha_{\pri}$ and $\alpha_{\prd}$ would imply a prior median of approximately 0.31 for $\pi_{\pri}$ and $\pi_{\prd}$ with a prior 95\% CI of (0.04, 0.80). This prior does not overly favor extreme parameter values, nor does it assign excess prior probability to one principal stratum over another. With covariates, the effective variance reduces by a factor approximately proportional to the number of covariate combinations (assuming roughly uniform distribution across covariate groups). To account for this, we thus increase the prior variance within each covariate by a factor of 4, i.e.~use a $N(0,2^2)$ prior for $\alpha_{g,\bm x}$. 
\item $\alpha_{\prh,\bm x} \sim N(-50,0.1^2)$. This prior ensures that $\pi_{\prh}=0$ with prior probability  essentially equal to 1. 
\item $\theta_{g,\bm x}(0)\sim N(\logit(0.3), 2^2)$, for all $g$, i.e.~identical and independent priors in all principal strata. The mean of this prior was chosen to reflect the expected two-year disability rate among untreated patients as described in the EXPAND study protocol. The variance was motivated with similar reasoning as above; in the absence of covariates, a $N(\logit(0.3), 1)$ prior would imply a prior median disability rate of 0.3 with a 95\% CI of (0.06, 0.75) which well covers the range of plausible values. Since we use two dichotomous covariates, the prior variance within each covariate combination is increased by a factor of 4. 
\item With $\theta_{g,\bm x}(1)$ parameterized as $\theta_{g,\bm x}(0) + \Delta_{g,\bm x}$, $\Delta_{g,\bm x}$ represents the difference from placebo (on a log-odds-ratio scale) when treated with active treatment. For $\Delta_{g,\bm x}$ we used a $N(0,2^2)$ prior to represent a prior expectation of no treatment effect, with the variance motivated exactly as above. 
\end{itemize}

Importantly, we avoided the use of flat non-informative priors (e.g.~$N(0,10^6)$) because such distributions can place a large amount of prior probability on regions of the parameter space that are highly implausible. For example, a diffuse normal prior for $\alpha_g$ would yield an implied bimodal prior for $\pi_g$ with most of the prior mass concentrated near 0 or 1. Similarly for $\theta_{g,\bm x}(0)$, even a moderately diffuse prior (e.g.~$N(0,50^2)$) implies only a 0.03 prior probability for the range 0.03 to 0.75 (which is the actual 95\% credible interval obtained with the prior specified above). 

To assess sensitivity to departures from monotonicity, we investigated two alternative prior distributions for $\alpha_{\prh, x}$. For ``weak monotonicity", we used a $N(-2,0.5^2)$ prior which implies (in the absence of covariates) a prior median of 0.04 for $\pi_{\prh}$ and a prior 95\% CI of (0.01, 0.13). In other words, this prior allows for the possibility that some patients might belong to the harmed stratum, but assigns low prior probability to this relative to the other principal strata. For ``no monotonicity", we used the same prior for $\alpha_{\prh,x}$ as for $\alpha_{\pri,x}$ and $\alpha_{\prd,x}$ (i.e.~$N(0,2^2)$). In this setting, the prior probability of belonging to the harmed stratum is not expected to be any larger or smaller than for the other strata. 

Finally, we note that because $\alpha_{\prb}=0$, the implied prior for $\pi_{\prb}$ is not identical to those of $\pi_{\pri}$ and $\pi_{\prd}$. Indeed, the prior median for $\pi_{\prb}$ is centered at 0.29 and the prior 95\% CI is somewhat narrower. Different priors could be used for $\alpha_{\pri}$ and $\alpha_{\prd}$ that would result in a more ``symmetric" prior distribution for $\pi_g$ ($\pi_{\prh}$ notwithstanding). For example, a bivariate normal prior for $(\alpha_{\pri},\alpha_{\prd})$ with a correlation of $0.5$ would result in approximately equal prior distributions. Working on the natural parameter scale and placing a Dirichlet$(1,1,1)$ prior on $(\pi_{\pri}, \pi_{\prd}, \pi_{\prb})$ would also achieve the same goal. Both of these prior configurations were investigated and did not produce results substantially different from those shown in Section \ref{sec:results}.

\subsection{Details on estimation}

\begin{table}
\centering
\caption{Summary statistics for the $t^*=12$ month time point, grouped by
anonymized covariate stratum and randomized treatment assignment. Statistics presented are:
number of patients randomized, number available for analysis, and number of relapses/CDPs observed between randomization and the given time point. \label{tab:data}}
\begin{tabular}{|rlrrrr|}\hline
Cov.~stratum & Treatment & \# randomized &
\# available & \# relapses & \# CDPs \\ \hline
1 & Siponimod & 208 & 167 & 22 & 30\\
1 & Placebo & 107 & 81 & 13 & 22\\
2 & Siponimod & 300 & 236 & 15 & 51\\
2 & Placebo & 155 & 126 & 17 & 35\\
3 & Siponimod & 180 & 145 & 20 & 20\\
3 & Placebo & 95 & 74 & 11 & 18\\
4 & Siponimod & 408 & 317 & 13 & 61\\
4 & Placebo & 188 & 137 & 7 & 20\\
 \hline
\end{tabular}
\end{table}

The model described in Section \ref{sec:model} was fitted using the probabilistic programming language Stan. This language provides Bayesian inference through Markov chain Monte Carlo (MCMC) methods using the No-U-Turn sampler (NUTS). See \cite{carpenter2017} for an overview. Four chains were simulated each with 1000 warm-up (tuning) iterations and 1000 sampling iterations which were saved for inference. Chains were randomly seeded, and mixing and convergence were assessed using graphical methods (e.g.~trace plots), diagnostics such as Rhat \citep[p.~285]{gelman2013bayesian} and checking for divergent transitions. Table \ref{tab:data} shows the summary statistics for the $t^*=12$ month time point, grouped by covariate stratum and treatment assignment. The Stan model file is provided in an online supplement. 

\subsection{Results}\label{sec:results}

\begin{figure}
\begin{center}
\includegraphics[height=4in]{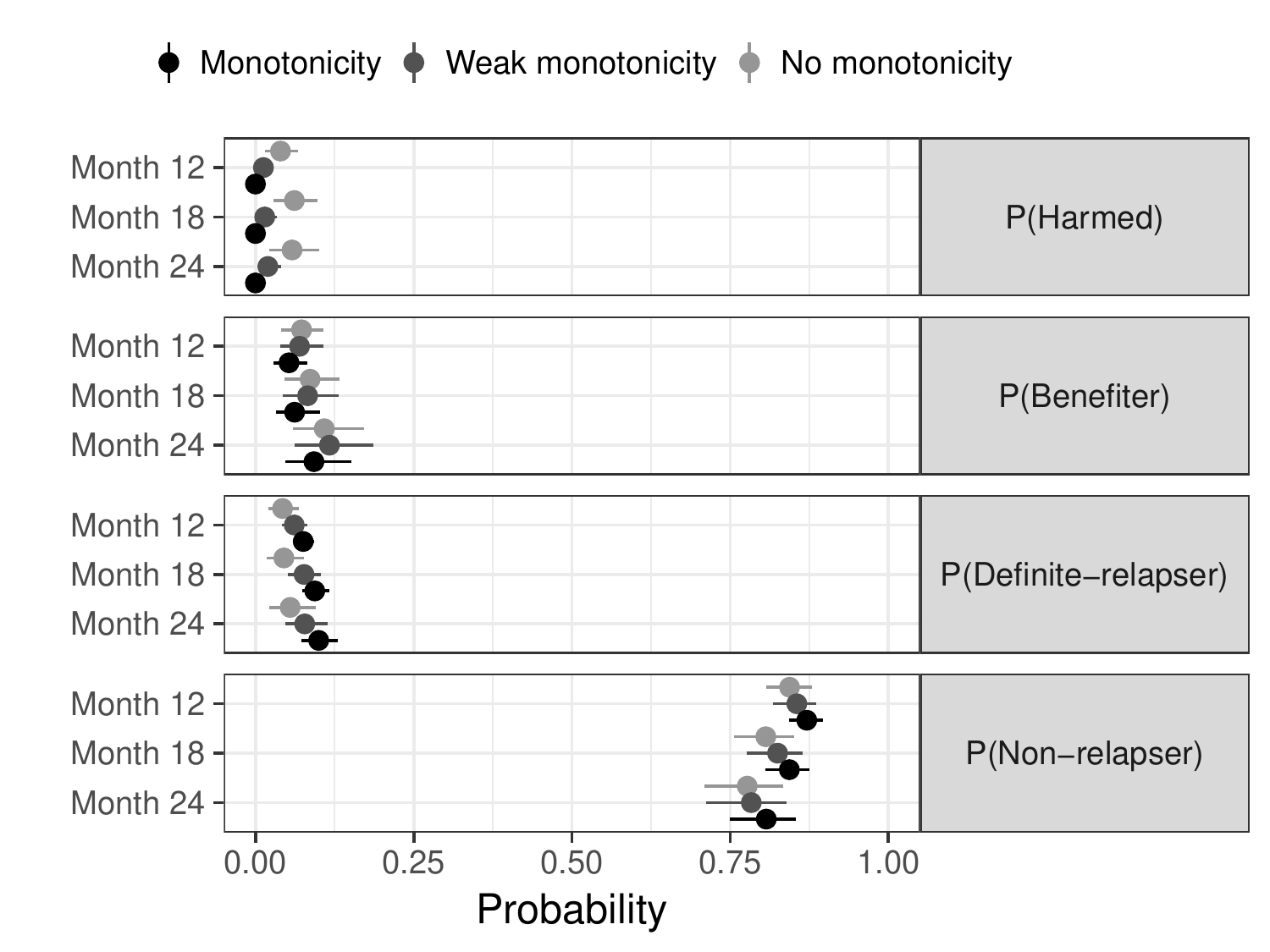}
\end{center}
\caption{Posterior probability of belonging to each principal stratum for different times $t^*$ (12, 18 and 24 months) and under varying assumptions for monotonicity. Shown are posterior medians with 95\% credible intervals. The ``Definite-relapser" and the ``Non-relapser" principal strata correspond to the ``doomed" and ``immune" principal strata, respectively, as discussed in Section \ref{sec:ps}.\label{fig:PS_prop}}
\end{figure} 

Estimated principal strata proportions are shown in Figure \ref{fig:PS_prop}. We see that the posterior probability of belonging to the non-relapser stratum is substantially larger than that of any other strata, with median probability of at least 0.8 under monotonocity. As the monotonicity assumption is relaxed, the posterior probability of belonging to the non-relapser remains the largest among all strata, and that for the harmed stratum is at most (under no monotonicity) roughly similar to the definite-relapser stratum. 

\begin{figure}
\begin{center}
\includegraphics[height=3.0in]{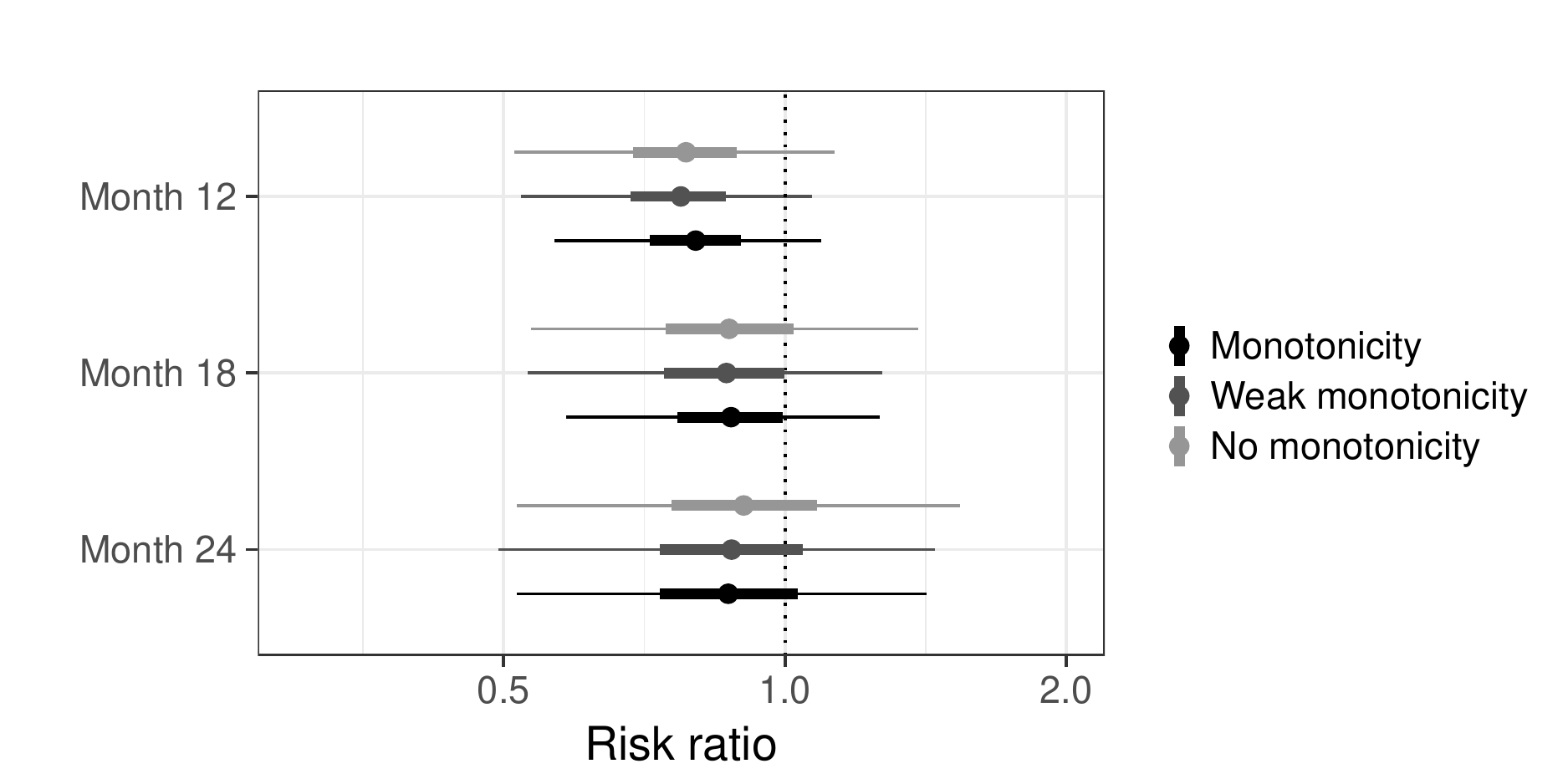}
\end{center}
\caption{Posterior risk ratios in the non-relapser principal stratum. Shown are posterior medians (point), and posterior 50\% and 95\% credible intervals (thick and thin lines, respectively). \label{fig:PS_rr1}}
\end{figure}

Figure \ref{fig:PS_rr1} shows inference for the estimand of interest (see Equation (\ref{eqn:estimand})). A risk ratio of $<1$ implies a beneficial effect of treatment. Under monotonicity we see a consistent benefit of treatment with posterior median risk ratios ranging from 0.80 to 0.85, depending on the time point. Further, there is at least approximately 70--75\% posterior probability that the risk ratio $<1$ for both endpoints and all time points. We note that the 95\% credible intervals get wider as $t^*$ increases. This is largely explained by decreasing number of patients available for analysis later in the trial (median follow-up time was approximately 18 months). 

The estimates remain fairly consistent as the monotonicity assumption is relaxed; posterior medians shift slightly but credible intervals are largely overlapping. While there is a slight increase in uncertainty (wider credible intervals) with weak or no monotonicity, it appears that the qualitative conclusions of the analysis do not depend strongly on this assumption. 

\section{Discussion} \label{sec:discuss} 

We have proposed a principal stratum estimand to quantify the effect of treatment in the population of patients that would not experience a post-randomization event based on potential outcomes. Our work is motivated by an interest in quantifying the efficacy of siponimod in a non-relapsing population of patients with secondary progressive multiple sclerosis. We used a Bayesian approach for statistical inference on the principal stratum estimand. This has the benefit of allowing some structural assumptions to be encoded using informative priors, which can be easily changed in sensitivity analyses. Furthermore, while the estimand of interest would be non-identifiable if estimated in the frequentist framework (see Section \ref{sec:identification}), working in the Bayesian framework allows straightforward calculation of the posterior distribution and derivation of inferential posterior summaries as long as sensible priors are used. Care is needed when specifying priors as results could be sensitive to particular choices. Very diffuse priors may be unintentionally informative in this context, and can also lead to computational difficulties. We used priors that capture the plausible range of parameter values. 
While not central to the methodology developed in this paper, we included covariates in the context of handling missing data. We used a missing at random assumption here, while a missing completely at random assumption was made in the companion article by \cite{cree2018}. Results for both missing data assumptions were comparable.   

The draft regulatory guidance document ICH E9(R1) on estimands and sensitivity analyses in clinical trials discusses five strategies (i.e.~treatment-policy, hypothetical, while on treatment, composite, principal stratum) for defining estimands in the presence of post-randomization events, referred to as ``intercurrent" events \citep{E9:2017}. In this article, we developed methods associated with the \textit{principal stratum} strategy in which the intercurrent event was the occurrence of relapse over a fixed time period and the outcome was disability progression over the same period. Alternatively, the \textit{treatment-policy} strategy would define an estimand that focuses solely on the occurrence of disability progression over a fixed time period and ignores the intercurrent event (i.e.~the intent-to-treat effect). This was the primary estimand in the EXPAND study. The \textit{composite} strategy would define a new variable that combines the intercurrent event and the outcome into a single variable, e.g.~no relapse and no disability progression over a fixed period of time. The estimand would then be the intention-to-treat effect based on the composite endpoint. The \textit{hypothetical} strategy would envisage a setting where relapses do not occur. This would require a precise description under which conditions this setting may be realistic. Finally, the \textit{while relapse-free} strategy (corresponding to the \textit{while on treatment} strategy in \cite{E9:2017}) would define a new variable, e.g. disability progression prior to relapse. 

The estimand framework has been discussed in the literature \citep{akacha2017a, leuchs2015, mallinckrodt2012, mehrotra2016}. Relatively speaking, the principal stratum strategy has received less attention. \cite{permutt2016} in his taxonomy of estimands highlights that the effect in the principal stratum is ``easy to define precisely, if not to estimate". \cite{akacha2017b} note that for the principal stratum strategy ``more formal training of clinical statisticians and practical experience is needed". The EXPAND study example described in this article provides such a practical example.

The methods developed in this paper apply to binary variables, both in the case of outcome and for intercurrent event. Extensions for continuously distributed outcome variables (e.g.~normally distributed) are straightforward. Another possible extension would be to treat both disability and relapse as time-to-event variables, for which one would define $S(z)$ and $Y(z)$ as the time to relapse and disability under treatment $z$, respectively. A patient would be then considered immune up to time $t$ if both $S(1) >t$ and $S(0) >t$. Survival distributions $F_{S(z)}(\cdot)$ and $F_{Y(z)}(\cdot)$ would need to be defined, for example with parametric or semi-parametric models. The joint survival distribution $F_{S(z),Y(z)}(\cdot,\cdot)$ could be modeled with a copula. Other approaches include \citep{ding2017}, who develop methodology that constructs weighted samples based on principal scores, defined as the conditional probabilities of the latent principal strata given covariates. Their analysis does not rely on any modeling assumptions on the outcome variable. However, identification of the principal causal effects relies on several assumptions, including exclusion-restriction (no effect on the intermediate variable implies zero effect on the outcome, see e.g.~\cite{angrist1996}), which would not be an appropriate assumption in our setting. 

In this article, we considered the use of principal stratification to assess the treatment effect in a subgroup who would not relapse regardless of treatment assignment. The methods developed can be applied, for binary outcomes, to draw inference about the causal effect among compliers \citep{little2015} and the causal effect among survivors \citep{rubin2006}. The former estimand is highly relevant in the context of non-inferiority or equivalence trials and the latter estimand in the context of trials in which functional outcomes may be truncated by death \citep{E9:2017}.

\bibliographystyle{Chicago}
\bibliography{bib}

\end{document}